# An Empirical Study on Transfer Learning for Privilege Review

Haozhen Zhao
Data & Technology
Ankura Consulting Group, LLC
Washington DC, USA
haozhen.zhao@ankura.com

Shi Ye
Data & Technology
Ankura Consulting Group, LLC
Washington DC, USA
shi.ye@ankura.com

Jingchao Yang
Data & Technology
Ankura Consulting Group, LLC
Washington DC, USA
jingchao.yang@ankura.com

*Abstract*— Protecting privileged communications and data from inadvertent disclosure is a paramount task in the US legal practice. Traditionally counsels rely on keyword searching and manual review to identify privileged documents in cases. As data volumes increase, this approach becomes less and less defensible in costs. Machine learning methods have been used in identifying privilege documents. Given the generalizable nature of privilege in legal cases, we hypothesize that transfer learning can capitalize knowledge learned from existing labeled data to identify privilege documents without requiring labeling new training data. In this paper, we study both traditional machine learning models and deep learning models based on BERT for privilege document classification tasks in legal document review, and we examine the effectiveness of transfer learning in privilege model on three real world datasets with privilege labels. Our results show that BERT model outperforms the industry standard logistic regression algorithm and transfer learning models can achieve decent performance on datasets in same or close domains.

*Keywords— privilege review, predictive coding, deep learning, BERT*

## I. Introduction

In the United States civil litigation process, a party has the right to withhold certain documents from court's production request on the basis that they contain privileged content involving communications between attorneys and clients with the purpose of obtaining legal advice. Protection of privileged information from inadvertent disclosure is a critical component of the US legal system [1]. To safeguard this information, parties are often engaged in expensive processes to review documents for production [2]. Review costs continue to rise as the number of documents that need to be reviewed in cases only gets larger and larger based on what we see in projects in the industry over the years. The high stakes involved in privilege review makes it the most expensive part of document review. The legal market has resorted to leveraging supervised machine learning approaches, or called predictive coding in legal community, to automate/semi-automate the document review process. Existing research shows that predictive coding has the potential to substantially reduce review costs [3,4].

The classic supervised machine learning approach requires significant amount of training examples to perform well in tasks. To address this bottleneck, transfer learning is a new learning paradigm that aims to leverage data from existing domains to train models that can generalize to the task at hand [5]. This is especially appealing to the privilege review task in that not only would it reduce the review costs and facilitate the document review workflow, but also that the concept of privilege may be generalizable across cases in a sense, as the definition of privileged circumstances are well defined in law.

In this paper, we empirically study three kinds of machine learning models in privilege prediction and investigate the transferability of models trained on different datasets using different machine learning methods. Main contributions of this paper include experimenting transferred models in predictive coding and conducting experiments with datasets with true labels instead on simulated labels.

The paper is organized as follows: Section II we review existing research related to machine learning for legal document review and transfer learning. Section III presents the research questions. Section IV describes the datasets used, the machine learning methods and experimental protocol, and evaluation measures. Section V discusses the results and Second VI concludes the papers with remarks on future work.

## II. Related Work

### A. Machine Learning for Privilege Review

Though cost pressure has driven the eDiscovery industry to adopt predictive coding solutions in cases, there is not many published research on using predict coding for privilege review. Gabriel et al. is among the first to discuss the use of predictive coding in finding privileged information in documents and showed that machine learning technology could bring significant improvement on efficiency and reliability to the privilege review workflow [6]. Gronvall et al. compared the traditional keyword search method and predictive coding approach in identifying privileged documents in cases and presented an approach to distinguish effective keywords from less efficient keywords [7]. Chhatwal et al. combined keyword search and Convolutional Neural Networks through training and predicting on potential privileged hits to identify privileged documents [8]. Oard et al. proposed a comprehensive framework called MINECORE (stands for "MINimizing the Expected COst of REview") which combines manual review and automatic classification to minimize both review costs and potential misclassification costs [9]. They experimented with a simulated data set and used surrogates for responsiveness and privilege labels.

*B. Transfer Learning in Deep Learning*

Over the past decade, as deep neural networks learning becomes the most popular machine learning paradigm, researchers discovered a common interesting phenomenon in the layers: first layers tend to learn features that can be generalized across domains and last layers are features more specific to the task at hand [10]. This observation paves the way to generalize very deep convolutional neural network image recognition model trained on very large image sets to image analytics applications [11]. Team in [12] has successfully taken the transfer learning approach to build an image clustering application on top of VGG16 for the legal domain. In 2018, Google made publicly available Bidirectional Encoder Representations from Transformers (BERT), a large transformers-based language representation model with 110 million parameters and pretrained on massive texts [13]. BERT immediately acquired much traction in both academia and industry due to its effectiveness in transferring knowledge learned in general domain to specific task, replicating the achievements we have seen in computer vision to natural language processing [14]. Mokrii et al. recently evaluated the transferability of BERT-based neural ranking models in question answering tasks [15]. Chalkidis et al. studied the performance of BERT in zero shot and few shot transfer learning problems [16].

III. RESEARCH QUESTIONS

This paper studies leveraging transfer learning in predictive coding for privilege review. Below are the research questions:

1. How do traditional machine learning methods such Logistic Regression (LR) perform comparing to BERT-based neural networks models in identifying privileged documents?

2. How effective are pretrained machine learning models based on traditional LR, document embeddings (Doc2Vec) [18] or BERT in predicting privileged documents in a zero shot settings, i.e. the models are trained on data fully outside of the evaluation set?

IV. EXPERIMENTS

We have conducted experiments on three datasets.

*A. Datasets*

The only publicly available evaluation dataset for privilege review is the one used in TREC 2010 Legal Track, which however has reliability issue [17]. Following our previous studies on predictive coding for privilege review [7][8], we experiment on three confidential and non-public datasets from real matters. Dataset A, and B are from two different matters related to the telecommunication industry. Dataset C is from a matter in the healthcare industry. All the documents were reviewed and coded on their privilege status by attorneys. Of the three datasets, we excluded documents with extracted text of over 1MB in size and filtered out duplicated documents. This results in three datasets of which the population size ranges from 202,235 to 566,475, and richness from 3.8% to 19.6% (Table I).

TABLE I. DATASETS STATISTICS

| Dataset | #Documents | #Privileged | #Not Privileged | Richness |
|---|---|---|---|---|
| A | 202,235 | 39,653 | 162,582 | 19.6% |
| B | 346,506 | 13,130 | 333,376 | 3.8% |
| C | 566,475 | 100,202 | 466,273 | 17.7% |

For the experiments carried out in this paper, we randomly sample a 30% subset from each of the three datasets and randomly split 70% for training and 30% for evaluation for the three subsets. Table II gives the privileged and not privileged document counts in both the training and testing splits for each dataset.

TABLE II. EXPERIMENTAL DATASET SPLITS

| Dataset | Split | #Privileged | #Not Privileged | #Total |
|---|---|---|---|---|
| A | Train | 8,340 | 34,129 | 42,469 |
| A | Test | 3,566 | 14,635 | 18,201 |
| B | Train | 2,689 | 70,077 | 72,776 |
| B | Test | 1,193 | 29,993 | 31,186 |
| C | Train | 21,110 | 97,849 | 118,959 |
| C | Test | 9,098 | 41,885 | 50,983 |

*B. Methodology*

In our experimental protocol, we employ three machine learning algorithms: Logistic Regression with TFIDF representation of the data (LR-TFIDF), Logistic Regression with Doc2Vec representation of the data (LR-Doc2Vec) and fully connected classifier layer on BERT output (FC-BERT). For each algorithm, we train a model on the each of training splits in Table II and evaluate the performance on all the testing splits.

LR-TFIDF: Logistic regression is the current state of the art text classification algorithm in the eDiscovery industry. It's also the most effective text classification algorithm according to empirical evaluations on real world datasets [4]. In our experiment, we use the liblinear implementation of the logistic regression as packaged in the Scikit Learn library. We use the default settings and use TFIDF term weighting scheme in the classification pipeline.

LR-Doc2Vec: Doc2Vec is distributed representation of documents based on word embeddings [18] that has the potential to encode semantic meanings of the corpus. Three Doc2Vec model are trained respectively using all the data of the datasets in Table I before the privilege experiments. We used 200 as the dimension of the embedding vectors, set a max vocabulary size as 500,000 and excluded terms appeared only once. The model is trained for 40 epochs. Training of Doc2Vec models is completely unsupervised, without using the privilege labels. Then in the experiments, we converted both the training documents and testing documents into embedding vectors using the pretrained Doc2Vec models, then use those vectors to train the LR model and run the prediction. We hypothesize that the transferability of models is encoded in the Doc2Vec models trained using the underlying data.

FC-BERT: We use the 'bert-base-uncased' model for the BERT setup and set dropout rate as 0.1, then take the top-level representation and feed into a dense layer with a sigmoid activation to produce the predicted labels. BERT is fine-tuned using the training datasets. Then the fine-tuned models are used to predict the labels of the testing set of each dataset. We hypothesize that BERT from general domain fine-tuned using in domain data, can be used to encode the knowledge of privilege in documents.

In addition to the modeling and prediction within each dataset, we also use models trained on one dataset's training set to predict the labels of other two dataset's test set. This is to investigate to what extend models can be transferred across domains or matters.

We use precision, recall, F1 score and precision at 75% recall (P75R) as the measures to evaluate the results of different methods. Precision is the percentage of truly privileged documents in all "predicted" privileged documents. Recall is the percentage of "predicted" truly privileged documents out of all the truly privileged documents in the testing set. Precision and recall only characterize one aspect of the model performance, F1 score is a balanced measure to characterize model performance in a whole, which is the harmonic mean of precision and recall. In eDiscovery practice, oftentimes parties negotiate a mutually acceptable recall level in the case, which is generally 75%. Therefore, precision at 75% recall is a commonly used measure of predictive coding model performance.

## V. RESULTS AND DISCUSSION

We first analyze the performance of the three methods within each dataset, with the purpose to examine if LR-Doc2Vec and FC-BERT outperform the industrial benchmark algorithm—LR-TFIDF.

TABLE III. PERFORMANCE OF DIFFERENT METHODS WITHIN THE SAME DATASET

| Dataset | Method | Precision | Recall | F1 | P@75C |
|---|---|---|---|---|---|
| A | LR-TFIDF | 87% | 67% | 76% | 82% |
| A | LR-Doc2Vec | 73% | 51% | 60% | 56% |
| A | FC-BERT | 82% | 82% | **82%** | 88% |
| B | LR-TFIDF | 70% | 9% | 16% | 17% |
| B | LR-Doc2Vec | 42% | 3% | 5% | 9% |
| B | FC-BERT | 55% | 21% | **30%** | 12% |
| C | LR-TFIDF | 85% | 66% | 74% | 79% |
| C | LR-Doc2Vec | 71% | 46% | 55% | 52% |
| C | FC-BERT | 83% | 74% | **78%** | 82% |

Table III shows the results of the different methods on the three datasets and Figure 1 plots the precision and recall of LR-TFIDF and FC-BERT (omitting LR-Doc2Vec for readability). Results show that LR-TFIDF achieved best precision, but FC-BERT model achieved best recall and generally outperforms LR-TFIDF in terms of F1 score. In terms of precision at 75% recall, FC-BERT outperforms LR-TFIDF in datasets A and C, except in the lower richness dataset B where LR-TFIDF outperforms FC-BERT. In all the cases, LR-Doc2Vec is less effective than LR-TFIDF and FC-BERT. Overall, FC-BERT method is the most effective classification algorithm for these datasets, slightly better in most of the balanced measures than LR-TFIDF with difference less than or around 5%.

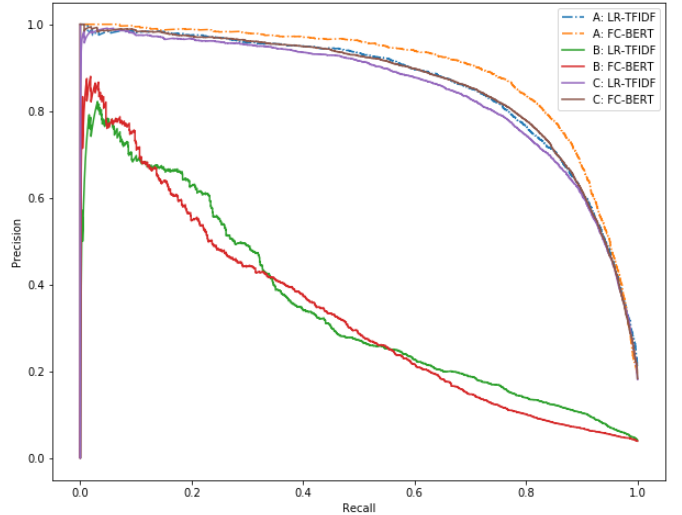

Fig. 1. Precision-Recall of LR-TFIDF and FC-BERT on Dataset A, B and C

We then study how transferable are the different models using each of the three datasets as the test datasets. We select the best model (according to F1 Score) from above as the benchmark in below comparisons.

TABLE IV. PERFORMANCE OF TRANSFERRED MODELS ON TEST DATA OF DATASET A

| Method | Train Dataset | Precision | Recall | F1 | P75R |
|---|---|---|---|---|---|
| Benchmark | A | 82% | 82% | **82%** | 88% |
| LR-TFIDF | B | 91% | 14% | 24% | 59% |
| LR-Doc2Vec | B | 77% | 4% | 8% | 44% |
| FC-BERT | B | 92% | 31% | <u>47%</u> | <u>65%</u> |
| LR-TFIDF | C | 88% | 8% | 15% | 44% |
| LR-Doc2Vec | C | 58% | 19% | 29% | 39% |
| FC-BERT | C | 64% | 23% | 34% | 38% |

TABLE V. PERFORMANCE OF TRANSFERRED MODELS ON TEST DATA OF DATASET B

| Method | Train Dataset | Precision | Recall | F1 | P75R |
|---|---|---|---|---|---|
| Benchmark | B | 55% | 21% | **30%** | 12% |
| LR-TFIDF | A | 54% | 9% | 15% | <u>11%</u> |
| LR-Doc2Vec | A | 28% | 16% | 21% | 8% |
| FC-BERT | A | 32% | 17% | <u>22%</u> | 8% |
| LR-TFIDF | C | 58% | 4% | 8% | 8% |
| LR-Doc2Vec | C | 23% | 10% | 14% | 7% |
| FC-BERT | C | 22% | 6% | 9% | 7% |

TABLE VI. PERFORMANCE OF TRANSFERRED MODELS ON TEST DATA OF DATASET C

| Method | Train Dataset | Precision | Recall | F1 | P75R |
|---|---|---|---|---|---|
| Benchmark | C | 83% | 74% | **78%** | 82% |
| LR-TFIDF | A | 77% | 0% | 1% | 26% |
| LR-Doc2Vec | A | 53% | 3% | <u>5%</u> | 26% |
| FC-BERT | A | 52% | 1% | 2% | 20% |
| LR-TFIDF | B | 82% | 0% | 0% | 19% |
| LR-Doc2Vec | B | 49% | 1% | 1% | <u>29%</u> |
| FC-BERT | B | 64% | 1% | 2% | 19% |

Table IV, V and VI show the results of transferring models on test data of new datasets. Figure 2 plots the precision and recall of the best performing zero shot transferred model of each dataset comparing with the benchmark model. For all three datasets, models trained on different datasets underperform the benchmark model which was trained on the test data from the same dataset, which is as expected. However, for Dataset A and Dataset B, transferred models perform quite well. For Dataset A the best performed transferred model is FC-BERT trained on Dataset B, which achieves 65% precision at 75% recall on A. For Dataset B, LR-TFIDF trained on A achieves 11% precision at 75% recall, close to the benchmark model's 12%.

For Dataset C, transferred models all perform badly, far behind the benchmark model. In this extreme case, LR-Doc2Vec surprisingly performs better than LR-TFIDF and FC-BERT.

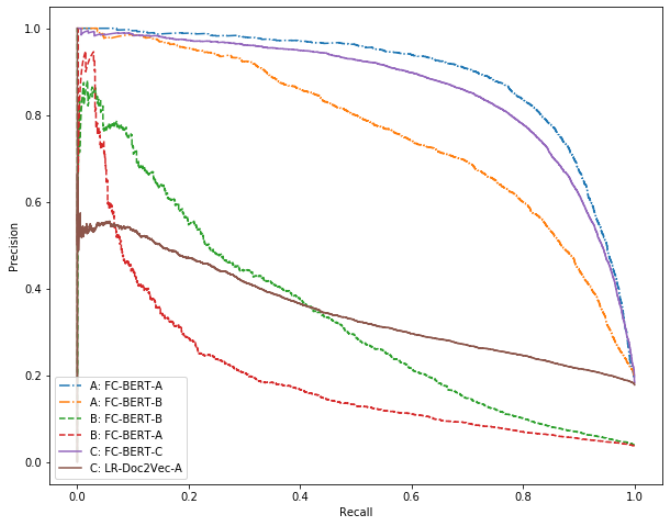

Fig. 2. Precision and Recall of best performing zero shot models and the bechmark models on Dataset A, B and C

One aspect of the results is that there seems no transitivity of transferred models. For example, models trained on training data of C perform much better on test data of A and B than that of models trained on training data of A or B on test data of C. This may be due to the size of the training data, as training set of Dataset C is larger than the training data of A or B.

The results show that domain plays a critical role in the transferability of models. Since Dataset A and Dataset B are from the same industry, i.e., telecommunications, models trained on one of them generally perform well on the other. As Dataset C is from a completely different industry, healthcare, models trained on A or B do not perform well on C.

## VI. CONCLUSIONS AND FUTURE WORK

This paper empirically studies the transferability of both traditional machine learning models and deep learning models on predicting privileged documents in legal document review. We have found that that BERT-based models with fine-tuning outperform logistic regression with TFIDF document representations according to F1 score. Zero shot transferred models though underperform natively trained models, can still achieve satisfactory performance if the target dataset is from a close domain.

The above conclusions are based on the utilized portion of the three datasets. In future, we plan to test with more data points and other available data sets. For the BERT-based models, we only add a single linear layer on top of BERT. It would also be worthwhile to explore with more sophisticated architecture, such as CNN or LSTM, etc. Also due to BERT restricting the input to only 512 Word Pieces, we currently have truncated long documents to only use the first 512 Word Pieces. More principled ways to deal with long documents in Transformers models will be explored in the next steps.